\documentclass{aa}  
\usepackage{graphicx,natbib,ulem}
\usepackage{txfonts}
\usepackage{color}
\begin{document}
  \titlerunning{Purple Mountain observations of SS 433}
\authorrunning{ Bowler}
   \title{On Purple Mountain H$\alpha$ observations of SS 433 }

   \subtitle{}

   \author{M. G.\ Bowler \inst{}}

   \offprints{M. G. Bowler \\   \email{m.bowler1@physics.ox.ac.uk}}
   \institute{University of Oxford, Department of Physics, Keble Road,
              Oxford, OX1 3RH, UK}
   \date{Received; accepted}

 
  \abstract
   {Certain  lines in spectra of the Galactic microquasar SS\,433, in particular  the brilliant Balmer H$\alpha$ line, have a two horned structure, interpreted as emission from a circumbinary disk orbiting a system of total mass approximately 40 $M_\odot$. It seems that the very existence of a circumbinary disk is regarded as controversial. Evidence bearing on the persistence and stability of these horns is therefor important.} 
  {To draw attention to the similarity between a set of H$\alpha$ spectra taken nightly over several orbital periods with a set of spectra taken at longer intervals, spread over several years; this similarity being relevant to persistence and stability of the circumbinary ring like features.}
   {Stationary H$\alpha$ spectra, taken almost
     nightly over two orbital periods of the binary system were analysed some years ago as superpositions of Gaussian components. Those spectra exhibit a two horned structure in excellent agreement with radiation from a circumbinary disk. A second set of H$\alpha$ spectra taken sporadically over a period of many years vary with orbital phase in much the same way. I have analysed this second set in terms of Gaussian components and quantified this resemblance. }  
  { The two sets of data are in remarkable agreement, despite the different timescales over which they were taken. This resemblance is even more marked when comparison is made for the common precession phases. The comparison also suggests that the tendency of the red horn to dominate the blue may be associated with obscuration by the wind from the accretion disk, the speed of which varies along the line of sight with precession phase.}
{ The twin horned structure of the stationary H$\alpha$ line is persistent and stable over many years and not just many days. This stability would seem to be natural if the origin is in a large scale circumbinary ring or disk; any alternative explanations must account for this persistence and stability.}


   \maketitle
%

\section{Introduction}

The Galactic microquasar  SS 433 is very luminous and unique in its continual ejection
of plasma in two opposite jets at approximately one quarter the speed
of light. The system is a 13 day binary and probably powered by supercritcal accretion by the compact member from its companion. The orbital speed of the compact object is fairly well established but in order to determine its mass either a measurement of the orbital velocity of the companion is needed or a measure of the total mass of the system. 

   Stationary emission lines in the spectra of SS 433 display a persistent two horn structure of just the kind expected for emission from an orbiting ring, or disk, seen more or less edge on. The horn separation corresponds to a rotation speed in excess of 200 km s$^{-1}$ and attributed to material orbiting the centre of mass of the binary system implies a system mass in excess of 40 $M_\odot$. The H$\alpha$ spectra were originally discussed in  Blundell et al (2008). Departures from the pattern expected for a uniformly radiating ring are present and are more pronounced in He I emission lines. These departures have been explained in terms of emission stimulated by some kind of spotlight rotating with the binary (Bowler 2010); propinquity of the intense radiation source in the accretion disk is sufficient (Bowler 2011).Thus these observations fix rather well the mass of the SS 433 system (hence of the compact object), provided that the two horned structure in H$\alpha$ and He I is indeed produced in an orbiting circumbinary ring rather than in ejecta on escape trajectories or some radially expanding structure. Origin in the inner rim of a circumbinary disk seems more likely to produce persistent and stable phenomena than the accretion disk blowing bubbles. The stability and persistence of the two horned structure is therefore of some significance.
   
 \begin{figure}[htbp]
\begin{center}
   \includegraphics[width=8cm]{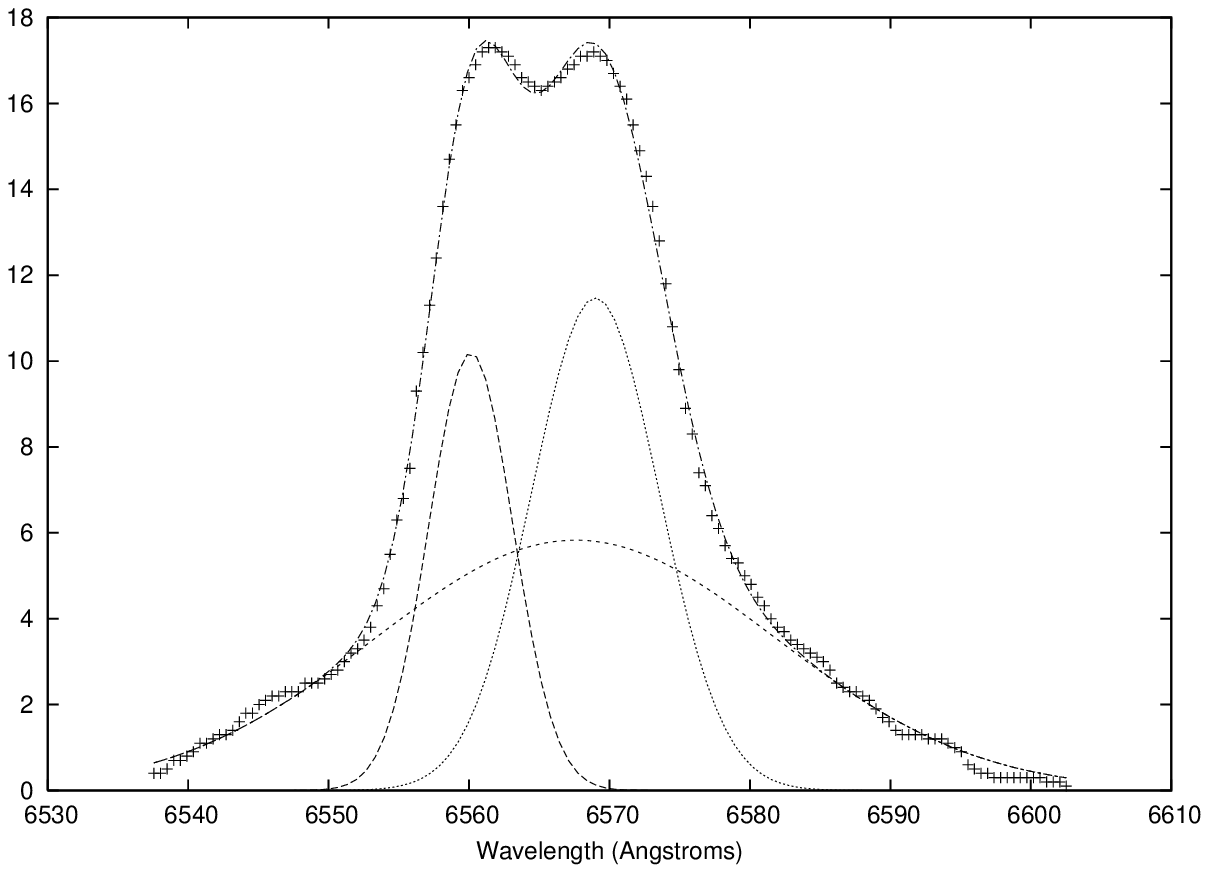} 
   \includegraphics[width=8cm]{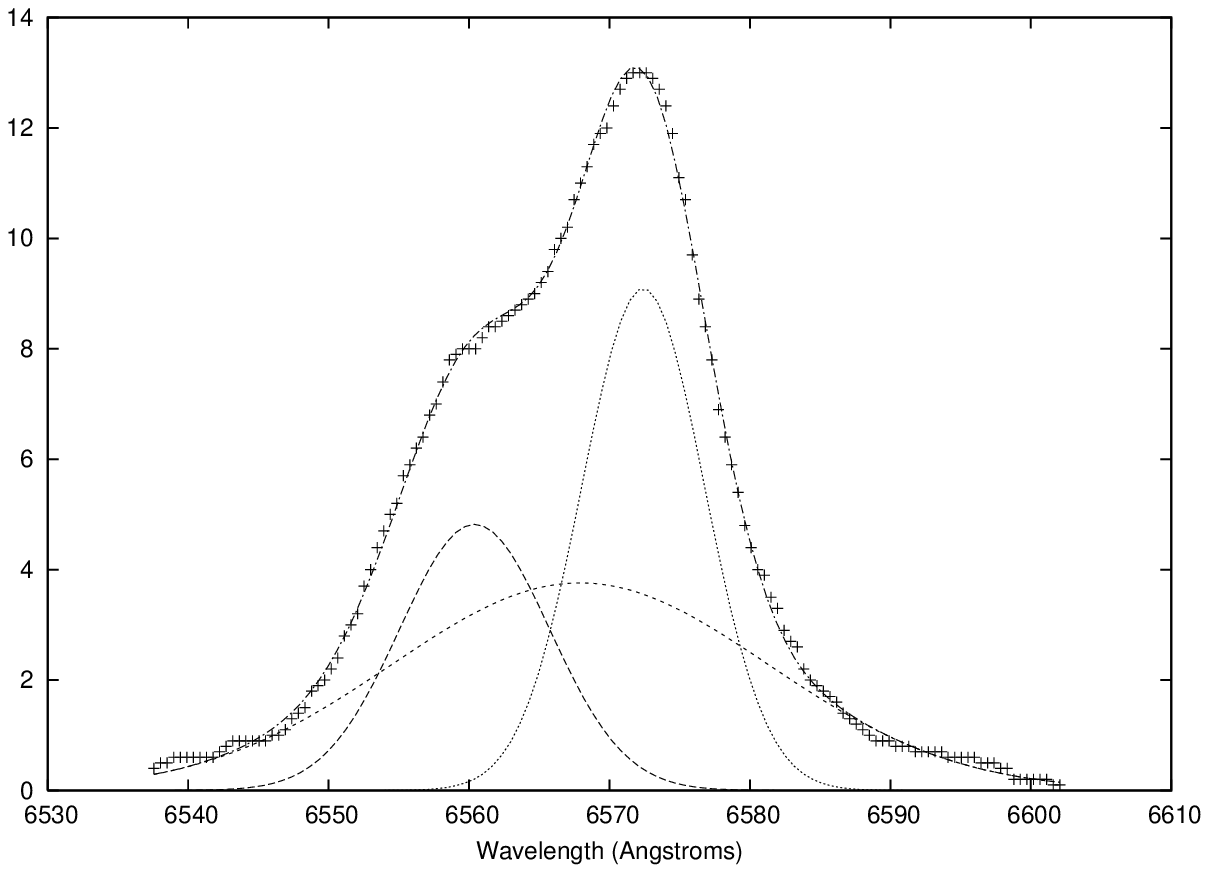}
\caption{ Two examples of the digitised Purple Mountain H$\alpha$ data and of its decomposition in terms of Gaussian components. The upper panel is for orbital phase 0.683 and the lower for orbital phase 0.212.
 }
\label{fig:disc}
\end{center}
\end{figure}

\section{The data sets compared}
\subsection{Qualitative comparisons}
  
  In this note I compare the SS 433 stationary H$\alpha$  spectra shown in Fig. 2 of Li \& Yan (2010) with the spectra in Fig. 2 of Schmidtobreick \& Blundell (2006a), as analysed in Blundell et al (2008) and Bowler (2010). These two sets of spectra were taken under very different conditions but are so similar in structure that it is my opinion that they show the same processes and that these are stable over a long period.
  The data of Li \& Yan (Purple Mountain Observatory) were taken in the years 2004, 2007 and 2008 and there are 17 spectra in all; labelled in that paper only according to the orbital phase. There is no other information provided except that the precessional phases lay in the range 0.155 and 0.296, where precessional phase 0 corresponds to the accretion disk most open towards the observer ( almost edge on for phase $\sim$0.3). In contrast, the ESO NTT data of Schmidtobreick \& Blundell (2006a,b) were taken over a single 30 day period in 2004, starting at JD 2453245 and ending JD 2453274. These data covered more than two orbits and the precessional phase varied from 0 at JD +245 to + 0.2 at +274. [There are also spectra taken almost every night between JD +287 and +310 but they do not concern me here because the H$\alpha$  line broadened in an optical flare preceding a radio flare. Two wings corresponding to red and blue shifted  velocities of more than  500 km s$^{-1}$ appeared and the signals I deal with here were confused - matters were made worse by P Cygni absorption troughs.]

A qualitative comparison of Fig.2 of Li \& Yan (2010) and Fig. 2 of Schmidtobreick \& Blundell (2006a) shows the stationary H$\alpha$  line to have, in both sets of data, a two horned structure on top of a broader component. In Schmidtobreick \& Blundell the bluer horn is dominant at JD +245 and the spectral shape evolves through a symmetric configuration (equal heights) at +248 to red horn dominant at about +253 (there was no observation on JD +252), just before orbital phase 0 at JD +255. The pattern of oscillation from blue horn strength to red horn dominance and back repeats throughout the 30 day period; the red becomes slowly more dominant (or equivalently blue is slowly more depleted than red) as time goes on. The Li \& Yan data show red dominance at orbital phase $\sim$ 0, evolving through a symmetric configuration at orbital phase $\sim$ 0.5 and returning to red dominance as orbital phase continues to 1. These two sets of data are qualitatively both consistent with the red and blue horns oscillating in antiphase and with blue progressively more strongly absorbed after a precessional phase of  $\sim$0.1. (The He I spectra in Schmidtobreick \& Blundell 2006a show very strong periodic swings from red to blue and back, throughout the 30 day period.)

In order to make quantitative comparisons I digitised the 17 spectra presented in Fig. 2 of Li \& Yan (2010) and fitted each to a sum of three Gaussians. The variables were the positions of the centroids, the standard deviations of the Gaussians and the heights at the centroids. The fits achieved a good representation of the spectra in all cases; the red and blue horns were generated by two Gaussians about 10 \AA\ apart and of standard deviation 3-4 \AA\ . These sit on a broader Gaussian of standard deviation approximately 13 \AA\ . Fig.1 displays two samples of my digitised Purple Mountain spectra, together with the fitted curves and the three Gaussian components.

 \subsection{Quantitative comparisons}
  
Fig.2 shows the centroids of the three fitted Gaussians as a function of orbital phase and is designed to be compared directly with Fig.1 of Blundell et al (2008) (which however covers over two orbital periods). The symbols $+$ and x denote the centroids of the blue and red horn Gaussians respectively; $*$ denotes the centroid of the broad component identified with the wind from the disk in Blundell et al (2008). The two narrow components run almost railroad straight and the wind centroid wanders from one side to the other with the orbital period, just as in Fig. 1 of Blundell et al (2008).
 
\begin{figure}[htbp]
\begin{center}
   \includegraphics[width=15cm,trim=0 0 0 2]{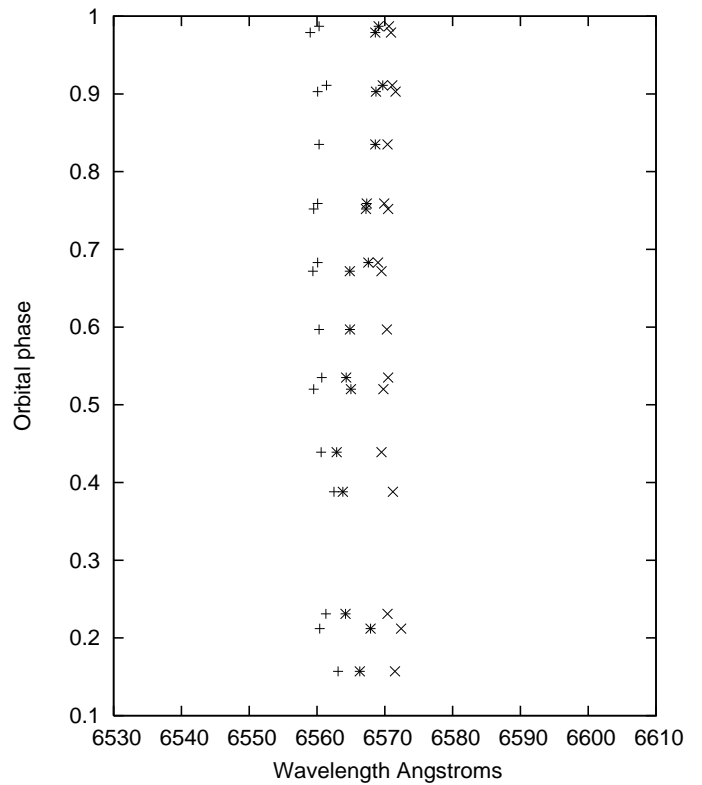} 
   \caption{ Doppler shifted wavelengths of the fitted Gaussian components of the Purple Mountain H$\alpha$ line. The blue component attributed to the circumbinary disk is denoted by $+$ and the red by x. The centre of the broad Gaussian from the accretion disk wind is denoted by $*$; it snakes between the narrower blue and red components. This figure is to be compared with Fig.1 of Blundell et al 2008.}
\label{fig:timesequence}
\end{center}
\end{figure}

Fig.3 presents the data of Fig.2 in a different way. The upper panel shows half the difference in recessional velocity of the two horn centroids; that is, approximately the rotational speed of the circumbinary ring which I have supposed responsible. This is to be compared with the upper panel of Fig.3 of Blundell et al (2008) or with the upper panel of Fig.7 of Bowler (2010). In both cases this supposed rotational speed is rather constant at approximately 200 km s$^{-1}$ (the Purple Mountain speed is maybe a bit larger than that from Chile). The middle panel is the complement; it is the mean speed of the horns and again can be compared with the upper panels of Fig. 3 of Blundell et al (2008) or Fig.7 of Bowler (2010). The nominal systemic speed of the horns looks a bit larger in the Purple Mountain data. The lowest panel displays the recessional speed of the centroid of the wind component ($*$ in Fig.2) as a function of orbital phase. This can be compared with the top panel of Fig.2 of Blundell et al (2008). In both data sets the wind centre is receding fastest at orbital phase $\sim$0.9 and approaching fastest at orbital phase $\sim$0.4. The amplitudes are consistent.
    
\begin{figure}[htbp]
\begin{center}
   \includegraphics[width=9cm,trim=0 0 0 120]{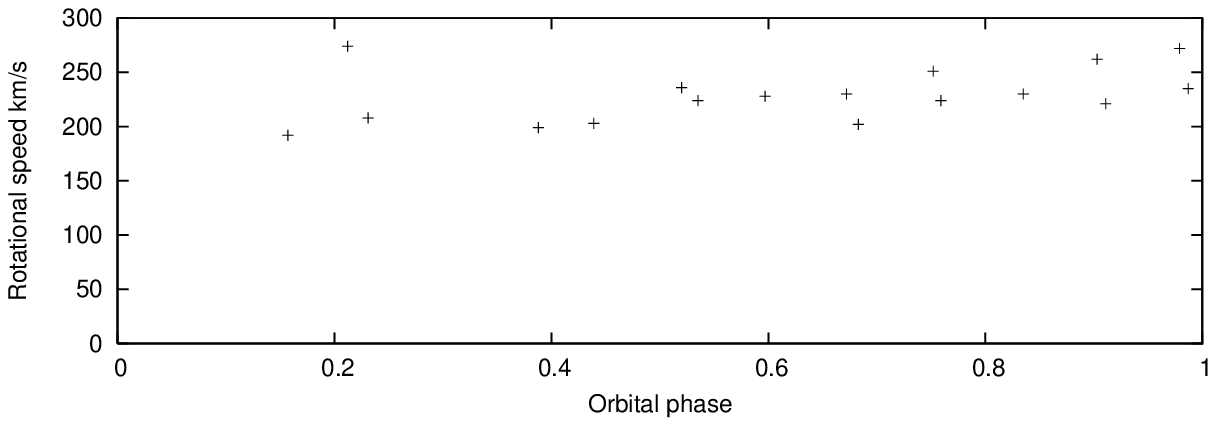}
   \includegraphics[width=9cm,trim=0 0 0 120]{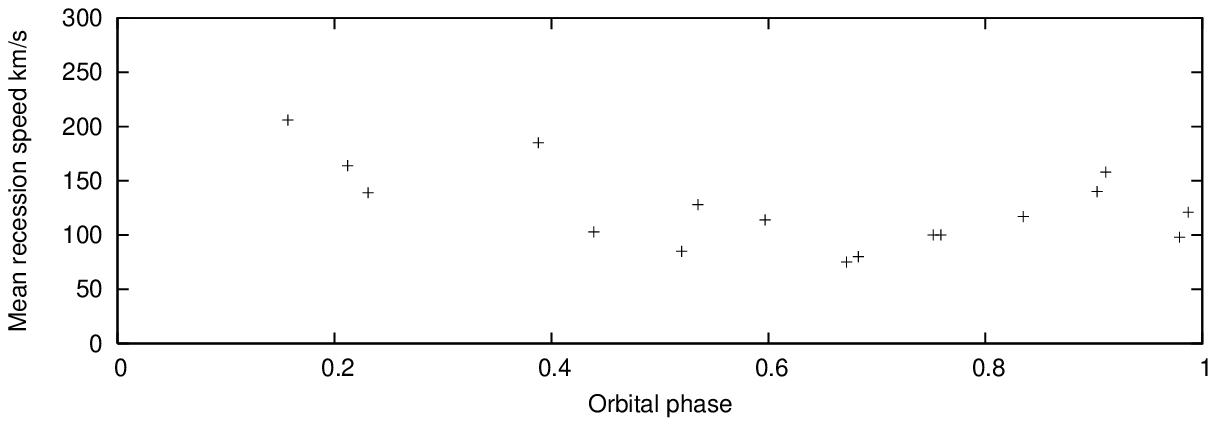}
   \includegraphics[width=9cm,trim=0 0 0 120]{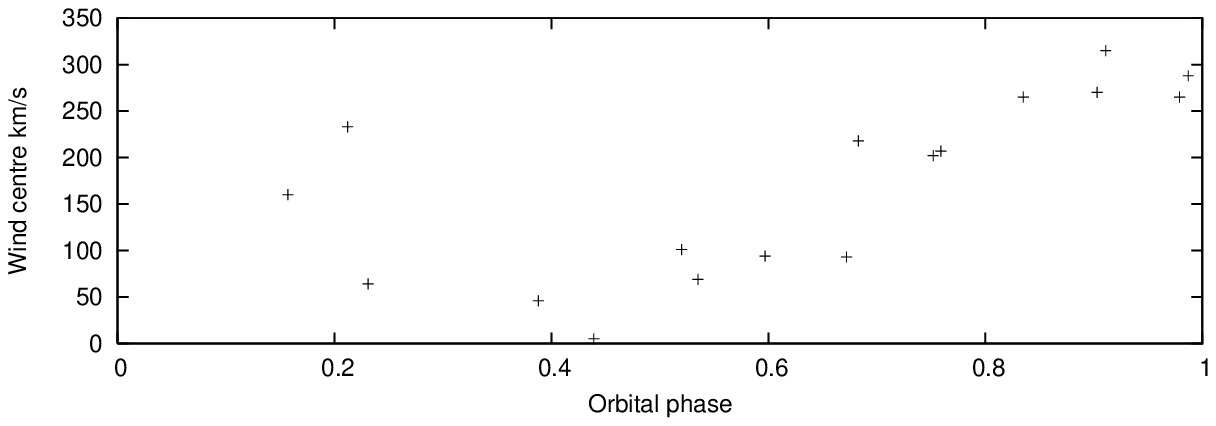}
   \caption{ The nominal rotational velocity of the circumbinary disk, as obtained from the differences between the red and blue components is shown in the top panel. The mean recessional speed of these components is in the middle panel and the variation of the recessional speed of the centre of the wind in the lowest panel, all displayed as a function of orbital phase.}
\label{fig:timesequence}
\end{center}
\end{figure}  

  
Finally, Fig.4 is concerned with the symmetry of the two horned structures. I have used an asymmetry parameter constructed in the following way: If the height of the red Gaussian is $h_r$   and of the blue $h_b$  then the asymmetry parameter is
  
\begin{equation}
 A=(h_r - h_b)/(h_r + h_b)
 \end{equation}

This quantity is plotted for the Li \& Yan data as a function of orbital phase in Fig.4 (it is roughly equivalent to the quantity R/V in Li \& Yan) in the upper panel. This asymmetry $A$ was not used in Blundell et al (2008) and so is plotted for the ESO data, covering more than two orbits, as a function of Julian Date in the lower panel. Orbital phase zero for those data occurs at JD +255. There is no real discrepancy and if differential absorption of the blue increases slowly with precessional phase then the agreement is very good.
  
\begin{figure}[htbp]
\begin{center}
   \includegraphics[width=9cm,trim=0 0 0 120]{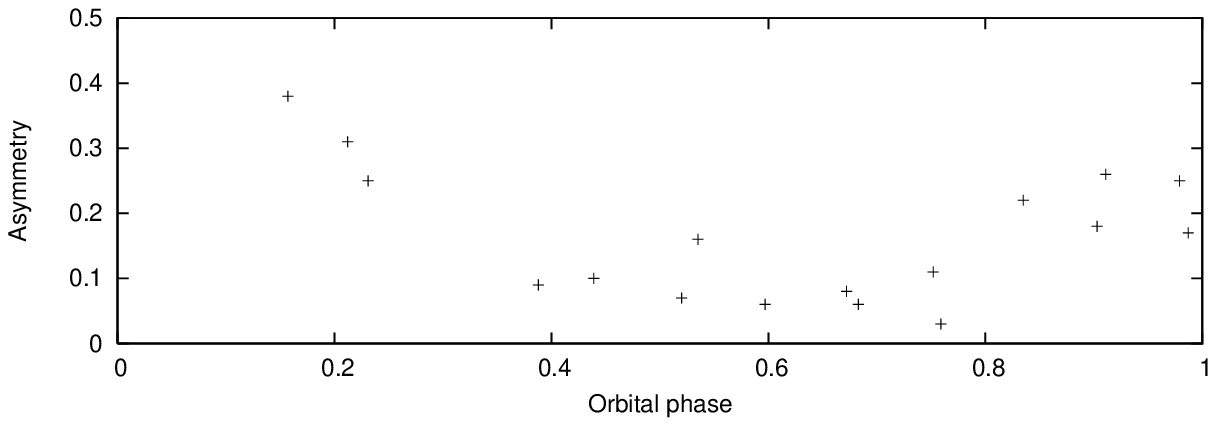}
   \includegraphics[width=9cm,trim=0 0 0 120]{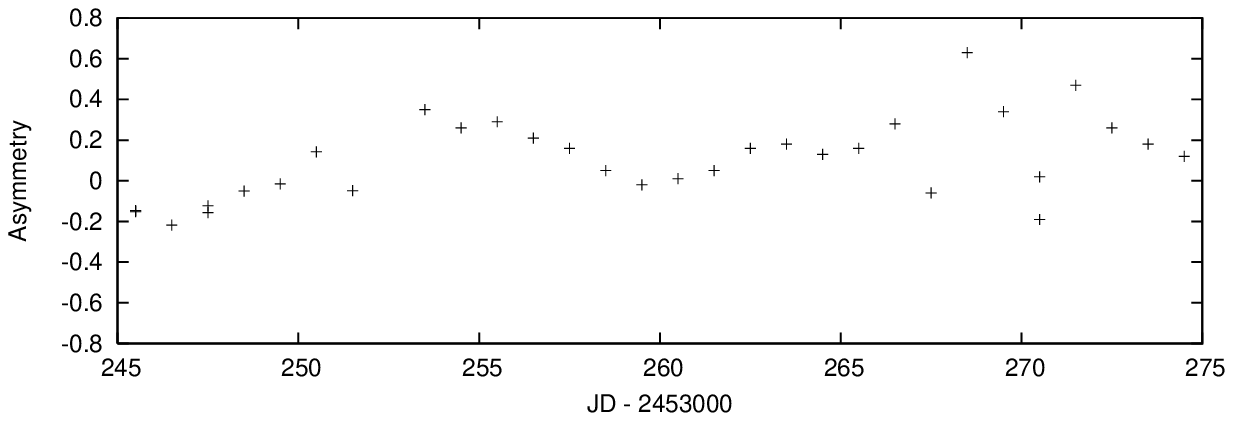}
   \caption{ The asymmetry of the red and blue circumbinary disk components as a function of time. The top panel is for the Purple Mountain H$\alpha$ data and the bottom panel is for the data of Blundell et al 2008. Note that the Purple Mountain data are plotted against orbital phase and the Blundell et al data as a function of Julian date. Orbital phase 0.5 occurs at JD +248 and orbital phase 0 at JD +255 in the latter.}
\label{fig:timesequence}
\end{center}
\end{figure}

  \section{Discussion}

Thus the spectra of Fig.2 of Li \& Yan (2010) and Schmidtobreick \& Blundell (2006a) seem to exhibit exactly the same underlying phenomena. These are (i) red and blue sources with standard deviations $\sim$ 3 \AA\, which oscillate in intensity in antiphase and which do not vary significantly in speed over the period of the orbit or indeed over a number of years. These have been attributed to a circumbinary disk (Blundell et al 2008, Bowler 2010,2011). Underlying these is (ii) an H$\alpha$  component formed in the wind from the accretion disk; this retains a memory of the orbital motion of the compact object and its disk (Blundell et al 2008). It looks like there is progressive differential reddening as the precession phase advances Ð I suggest that this may be because the blue is preferentially absorbed in winds or outflows from the system at about 200 km s$^{-1}$. Such absorption certainly occurs and at its most extreme is manifest in P Cygni troughs on the blue side of Balmer series lines - pretty much where the blue horn occurs at 6560 \AA\ .
    
    It should be noted that the data shown in Fig.2 of Li \& Yan (2010) have been subjected to some selection. In Fig.1 of that paper are shown two examples of spectra taken at precessional phases outside of the range 0.155 to 0.296. One spectrum taken at precessional phase close to 0 (accretion disk wide open to the observer) seems to show a single dominant peak; a second taken at precessional phase 0.7 has high velocity wings, of much the same kind as observed after JD + 287 in Fig.2 of Schmidtobreick \& Blundell (2006a) and associated with a flare. These intermittent high speed components might correspond to unveiling of inner regions of the accretion disk, or possibly to ejections from its vicinity; in either case the lines attributed to the circumbinary disk are confused or swamped to various degrees for the duration of the outburst.

\section{Conclusions}

The agreement between the Purple Mountain spectra, taken at various times in 2004, 2007 and 2008, with the Chile data taken over a single 30 day period in late 2004 seems a powerful argument for the red and blue horns being a permanent feature of the H$\alpha$  spectrum. The two horned structure was seen in various spectral lines by Filippenko et al (1988), in H$\alpha$  and He I by Gies et al (2002) and in H$\alpha$ and other spectral lines by Schmidtobreick \& Blundell (2006a,b). Thus the two horn phenomenon persists over long periods and the speed attributed to a ring source is extremely stable. The Purple Mountain data are of great importance in establishing this stability and the stability is important because it argues for an origin in a circumbinary ring, as proposed by Blundell et al (2008). I find it hard to imagine that a source in radially expanding rings blown off the accretion disk of SS 433 or ring structures formed somehow in the wind from the disk could exhibit this degree of stability, both short and long term. (D. R. Gies, in a private communication, gave the opinion that such an expanding ring is a much more plausible origin for the two horned structure than a circumbinary disk.) If the source is a circumbinary ring orbiting the system, then the system mass exceeds 40 $M_\odot$ and the compact object is a rather massive stellar black hole.

\begin{acknowledgements}
I acknowledge useful correspondence with J. Z. Yan.

\end{acknowledgements}

\end{document}